\documentclass[12pt,a4paper,twoside]{article}
\newenvironment{changemargin}[2]{%
\begin{list}{}{%
\setlength{\leftmargin}{#1}%
\setlength{\rightmargin}{#2}%
}%
\item[]}
{\end{list}}
\usepackage{graphicx}
\usepackage{color}
\usepackage[
top    = 2.75cm,
bottom = 2.50cm,
left   = 1.50cm,
right  = 1.50cm]{geometry}
\usepackage{epstopdf}
\usepackage{pdflscape}
\usepackage{amsmath}
\usepackage{amsmath}
\usepackage{amsmath}
\usepackage{kantlipsum}
\allowdisplaybreaks
\usepackage{setspace}

\usepackage{array}

\newcolumntype{M}[1]{>{\raggedright\arraybackslash}m{#1}}
\newcolumntype{N}{@{}m{0pt}@{}}

\begin{document}
\baselineskip=0.25in
{\bf \LARGE
\begin{changemargin}{-0.5cm}{-0.5cm}
\begin{center}
{JRSP of three-particle state via three tripartite GHZ class in quantum noisy channels}
\end{center}\end{changemargin}}
\begin{center}
\textit{Int. J. Quantum Inform. {\bf14} (2016) 1650034}
\end{center}
\vspace{4mm}
\begin{center}
\large{\bf Babatunde James Falaye $^{a,b,}$}\footnote{\scriptsize E-mail:~ fbjames11@physicist.net; babatunde.falaye@fulafia.edu.ng}\large{\bf ,} {\large{\bf Guo-Hua Sun $^{c}$}}\footnote{\scriptsize E-mail:~ sunghdb@yahoo.com}\large{\bf ,} {\large{\bf Oscar Camacho-Nieto $^{d}$}}\footnote{\scriptsize E-mail:~ ocamacho@ipn.mx} \large{\bf and} {\large{\bf Shi-Hai Dong $^{d}$}}\footnote{\scriptsize E-mail:~ dongsh2@yahoo.com }
\end{center}
{\footnotesize
\begin{center}
{\it $^\textbf{a}$Departamento de F\'isica, Escuela Superior de F\'isica y Matem\'aticas, Instituto Polit\'ecnico Nacional, Edificio 9, UPALM, M\'{e}xico D.F. 07738, M\'{e}xico.}\\
{\it $^\textbf{b}$Applied Theoretical Physics Division, Department of Physics, Federal University Lafia,  P. M. B. 146, Lafia, Nigeria.}\\
{\it $^\textbf{c}$C\'{a}tedr\'{a}tica CONACyT, CIC, Instituto Polit\'{e}cnico Nacional, UPALM, M\'{e}xico D. F. 07700, M\'{e}xico.}\\
{\it $^\textbf{d}$CIDETEC, Instituto Polit\'{e}cnico Nacional, UPALM, M\'{e}xico D. F. 07700, M\'{e}xico.}
\end{center}}
\begin{center}

\end{center}
\begin{abstract}
\noindent
We present a scheme for joint remote state preparation (JRSP) of three-particle state via three tripartite Greenberger-Horne-Zeilinger (GHZ) entangled states  as the quantum channel linking the parties. We use eight-qubit mutually orthogonal basis vector as measurement point of departure. The likelihood of success for this scheme has been found to be $1/8$. However, by putting some special cases into consideration, the chances can be ameliorated to $1/4$ and $1$. The effects of amplitude-damping noise, phase-damping noise and depolarizing noise on this scheme have been scrutinized and the analytical derivations of fidelities for the quantum noisy channels have been presented. We found that for $0.55\leq\eta\leq1$, the states conveyed through depolarizing channel lose more information than phase-damping channel while the information loss through amplitude damping channel is most minimal.
\noindent
\end{abstract}


Depolarizing noise.

{\bf PACs No.}: 03.65.Ud, 03.67.Hk, 03.67.Ac, 03.67.Mn.

\section{Introduction}
\label{sec1}
When a pair of particles is generated such that the quantum state of each particle cannot be elucidated independently, we referred to them as being entangled. One of the effortless ways to create entanglement is using a very powerful laser along with some very special crystals in order to entangle pair of photons which are the smallest unit of light. Experimental realization of quantum entanglement has been achieved and presented more elaborately in reference \cite{T1}. Quantum entanglement represents basic ingredient in quantum information processing. In fact, quantum entanglement lies in the heart of many quantum devices that are actively been developed. Entanglement has been found resourceful in quantum state sharing \cite{A1}. Information stored in quantum system can be teleported from one location to the other with the aid of shared entangled state \cite{A2}. Some other processes that exploit quantum entanglement include quantum dense coding \cite{A3}, quantum secure direct communication \cite{A4} and remote state preparation (RSP) \cite{A5}.

The quantum teleportation can be described as a process by which quantum information is being transmitted from one location to another, with the aid of classical communication and erstwhile shared quantum entanglement between the sending and receiving location. In quantum teleportation, the sender has a particle of {\it unknown state}. Whereas, in RSP, the sender does not own the particle but all the classical information about the state of the particle to be prepared for a receiver, who is remotely sundered from the sender. However, like the quantum teleportation, a shared entangled state is a prerequisite as quantum channel in RSP. With respect to this, RSP can be regarded as teleportation of {\it known state}. This idea, which was  expounded independently by Lo, Patil and Bennett et al. \cite{A5}, shows that the communication cost is lower than that of teleportation protocol \cite{A2}. In RSP scheme, the sender, say Alice performs a projective measurement on her qubits in the shared entangled state with the receiver, say Bob, and then communicates the result to Bob via classical channel. Depending on the outcome of the measurement, Bob can apply an appropriate quantum gate to reconstruct the original particle's state that Alice intends to transfer from the shared entangled state.

After the first proposed RSP protocol, there have been several proposed RSP protocols from many researchers (\cite{A7, A8} and references therein). Remote simulation of any single-particle measurement on an arbitrary qubit using one ebit of shared entanglement and communication of one cbit, has been achieved in reference \cite{TR3}. Furthermore, experimentally, RSP has been realized in reference \cite{A9} and via noisy entanglement in reference \cite{A10}. RSP of photonic hybrid from spin to orbital angular momentum degrees of freedom has been recently realized experimentally in reference \cite{TR2}. Peters et al. \cite{TR4} reported experimental demonstration of RSP of arbitrary single-qubit state, conceal in the polarization of photons which are generated by spontaneous parametric down-conversion. Moreover, it is worth mentioning that due to all information being apportioned to one sender, there could be leakage in information (suppose Alice is not sincere) which could lead to precarious protocol. To overcome this challenge, Xia et al. \cite{A11} proposed joint remote state preparation (JRSP) where there are two parties to prepare the state for the remote receiver.

In JRSP, information of the state to be prepared is shared by two or more senders situated at different locations and consequently, no individual has the complete information. After the first proposed JRSP protocol by Xia et al., there have been several proposed probabilistic and deterministic JSRP protocols from many researchers (see \cite{A12,A13,A14} and references therein). The authors of reference \cite{TR5} present quantum circuits and photon circuits for jointly preparing one qubit state. Very recently An and Bich \cite{TR6}, constructed a quantum circuit to  fabricate a task-oriented partially entangled state and used it as quantum channel for controlled JRSP. Chen and Xia \cite{TR7} studied JRSP of an arbitrary two-qubit state using a generalized seven-qubit brown state as a quantum channel. So many outstanding experimental and theoretical researches have been reported so far which made it impossible to list all the contributions. However, the aforementioned references, form basic foundation for which the current work is built on.

In line with these perpetual interests, our contributions are in two folds. Firstly, the current study  extends the work of Wang \cite{A7} to tripartite system. Secondly, we study the effects of noisy environments on the communication protocol. JRSP of three-particle system was studied concisely in reference \cite{RT1}, consequently, the first fold of the current work may be considered as an extensive review of reference \cite{RT1}. To the best of our knowledge, JRSP of three-particle in quantum noisy channels has not been expounded till this moment which we fell it might be due to complexity in mathematical involved. Most efforts in literature were based on two-particle system (\cite{A13, A14} and references therein) whereas interaction of JRSP of three-particle system with environment is also not avoidable. It is therefore second fold and the priority objective of the current work to study JRSP of three-particle system in quantum noisy channels.

Three quantum noisy channels namely: amplitude-damping, phase-damping and depolarizing channels will be considered in this paper. Thus, using the fidelity, we shall determine information loss as quantum information is been transmitted through these noisy channels. Generally speaking, this study can also be considered as furtherance of some recent works \cite{A13, A14} where JRSP of two-qubit system have been studied in a noisy channel. In this paper, we shall employ three tripartite GHZ class \cite{A15,A16}; ${1}/{\sqrt{2}}\left(\left|000\right\rangle+\left|111\right\rangle\right)$ as quantum channel linking the three parties.

GHZ is a type of quantum  entanglement which involves at least three subsystems. GHZ is fully separable after loss of one qubit unlike W-class which is still entangled with remaining two-qubit. GHZ of three photons and three Rydberg atoms have been observed experimentally \cite{A17}. Using spontaneous parametric down-conversion, three-photon polarization-entangled W state has also been realized experimentally in reference \cite{A18}. The motivation behind GHZ experiment is due to the fact that GHZ states manifest strong quantum correlations, such that an elegant test of the nonlocality of quantum mechanics is possible \cite{TR8}.

This paper is organized as follows: In section $2$, we review a scheme for JRSP of three-particle entangled state using three tripartite GHZ class. In section $3$, we study JRSP subjected to amplitude-damping, phase-damping and depolarizing noises. Concluding remarks are given in section $4$.

\section{JRSP of an arbitrary three-particle state}
In this section, we review a protocol for JRSP involving two senders and one receiver in a closed quantum system. Now, let Alice and Bob be the two senders, who are located at spatially separated nodes, wish to help the receiver Chika in remote preparation of an arbitrary three-particle state which can be written as \cite{A7,RT1}
\begin{eqnarray}
\left|\Omega\right\rangle&=&\alpha_1e^{i\phi_1}\left|000\right\rangle+\alpha_2e^{i\phi_2}\left|001\right\rangle+\alpha_3e^{i\phi_3}\left|010\right\rangle+\alpha_4e^{i\phi_4}\left|011\right\rangle\nonumber\\
&+&\alpha_5e^{i\phi_5}\left|100\right\rangle+\alpha_6e^{i\phi_6}\left|101\right\rangle+\alpha_7e^{i\phi_7}\left|110\right\rangle+\alpha_8e^{i\phi_8}\left|111\right\rangle,
\end{eqnarray}
where the real coefficients $\alpha_i$,  $(i=1-8)$ satiate the normalization condition $\sum_{n=1}^8(\alpha_i)^2=1$ and $\phi_i\in [0, 2\pi]$. The information shared by Bob and Alice has been denoted by $\left|\Omega\right\rangle$. The phase information $\phi_i$ belongs to Bob while the amplitude information $\alpha_i$ is known to Alice. The quantum communication channel linking Alice, Bob and Chika can be written as:
\begin{eqnarray}
\left|\mathcal{F}\right\rangle_{123456789}=\frac{1}{2\sqrt{2}}\left(\left|000\right\rangle+\left|111\right\rangle\right)\otimes\left(\left|000\right\rangle+\left|111\right\rangle\right)\otimes\left(\left|000\right\rangle+\left|111\right\rangle\right),
\end{eqnarray}
where particles trios $(1,4,7)$, $(2,5,8)$ and $(3,6,9)$ belong to Alice, Bob and Chika respectively. Now Alice and Bob perform projective measurements on their respective particles trios $(1,4,7)$ and $(2,5,8)$ in order to remotely prepare original state for Chika. For these measurements to be achieved, Alice chooses a set of mutually orthogonal basic vector $\{\left|\varrho_{147}\right\rangle_n, n=1-8\}$ which are related to computational basis vector $\{\left|000\right\rangle, \left|001\right\rangle, \left|010\right\rangle, \left|011\right\rangle, \left|100\right\rangle, \left|101\right\rangle, \left|110\right\rangle, \left|111\right\rangle\}$ as follows
\begin{eqnarray}
\left(\begin{matrix}\left|\varrho_{147}\right\rangle_1\\ \left|\varrho_{147}\right\rangle_2\\ \left|\varrho_{147}\right\rangle_3\\ \left|\varrho_{147}\right\rangle_4\\ \left|\varrho_{147}\right\rangle_5\\ \left|\varrho_{147}\right\rangle_6\\ \left|\varrho_{147}\right\rangle_7\\ \left|\varrho_{147}\right\rangle_8\end{matrix}\right)=
\left(
\begin{matrix}
\alpha_1&  \alpha_2&  \alpha_3&  \alpha_4&  \alpha_5&  \alpha_6&  \alpha_7&  \alpha_8\\
\alpha_1& -\alpha_2&  \alpha_3& -\alpha_4&  \alpha_5& -\alpha_6&  \alpha_7& -\alpha_8\\
\alpha_1& -\alpha_2& -\alpha_3&  \alpha_4& -\alpha_5&  \alpha_6&  \alpha_7& -\alpha_8\\
\alpha_1&  \alpha_2& -\alpha_3& -\alpha_4&  \alpha_5&  \alpha_6& -\alpha_7& -\alpha_8\\
\alpha_1& -\alpha_2&  \alpha_3& -\alpha_4& -\alpha_5&  \alpha_6& -\alpha_7&  \alpha_8\\
\alpha_1&  \alpha_2& -\alpha_3& -\alpha_4& -\alpha_5& -\alpha_6&  \alpha_7&  \alpha_8\\
\alpha_1& -\alpha_2& -\alpha_3&  \alpha_4&  \alpha_5& -\alpha_6& -\alpha_7&  \alpha_8\\
\alpha_1&  \alpha_2&  \alpha_3&  \alpha_4& -\alpha_5& -\alpha_6& -\alpha_7& -\alpha_8
\end{matrix}\right)\left(\begin{matrix}\left|000\right\rangle\\ \left|001\right\rangle\\ \left|010\right\rangle\\ \left|011\right\rangle\\ \left|100\right\rangle\\ \left|101\right\rangle\\ \left|110\right\rangle\\ \left|111\right\rangle\end{matrix}\right).
\end{eqnarray}
Bob chooses $\{\left|\varsigma_{258}\right\rangle_n, n=1-8\}$ as his measurements basis. These are related to computational basis vector $\{\left|000\right\rangle, \left|001\right\rangle, \left|010\right\rangle, \left|011\right\rangle, \left|100\right\rangle, \left|101\right\rangle, \left|110\right\rangle, \left|111\right\rangle\}$ as follows
\begin{eqnarray}
\left(\begin{matrix}\left|\varsigma_{258}\right\rangle_1\\ \left|\varsigma_{258}\right\rangle_2\\ \left|\varsigma_{258}\right\rangle_3\\ \left|\varsigma_{258}\right\rangle_4\\ \left|\varsigma_{258}\right\rangle_5\\ \left|\varsigma_{258}\right\rangle_6\\ \left|\varsigma_{258}\right\rangle_7\\ \left|\varsigma_{258}\right\rangle_8\end{matrix}\right)=
\frac{1}{2\sqrt{2}}\left(
\begin{matrix}
e^{-i\phi_1}&  e^{-i\phi_2}&  e^{-i\phi_3}&  e^{-i\phi_4}&  e^{-i\phi_5}&  e^{-i\phi_6}&  e^{-i\phi_7}&  e^{-i\phi_8}\\
e^{-i\phi_1}& -e^{-i\phi_2}&  e^{-i\phi_3}& -e^{-i\phi_4}&  e^{-i\phi_5}& -e^{-i\phi_6}&  e^{-i\phi_7}& -e^{-i\phi_8}\\
e^{-i\phi_1}& -e^{-i\phi_2}& -e^{-i\phi_3}&  e^{-i\phi_4}& -e^{-i\phi_5}&  e^{-i\phi_6}&  e^{-i\phi_7}& -e^{-i\phi_8}\\
e^{-i\phi_1}&  e^{-i\phi_2}& -e^{-i\phi_3}& -e^{-i\phi_4}&  e^{-i\phi_5}&  e^{-i\phi_6}& -e^{-i\phi_7}& -e^{-i\phi_8}\\
e^{-i\phi_1}& -e^{-i\phi_2}&  e^{-i\phi_3}& -e^{-i\phi_4}& -e^{-i\phi_5}&  e^{-i\phi_6}& -e^{-i\phi_7}&  e^{-i\phi_8}\\
e^{-i\phi_1}&  e^{-i\phi_2}& -e^{-i\phi_3}& -e^{-i\phi_4}& -e^{-i\phi_5}& -e^{-i\phi_6}&  e^{-i\phi_7}&  e^{-i\phi_8}\\
e^{-i\phi_1}& -e^{-i\phi_2}& -e^{-i\phi_3}&  e^{-i\phi_4}&  e^{-i\phi_5}& -e^{-i\phi_6}& -e^{-i\phi_7}&  e^{-i\phi_8}\\
e^{-i\phi_1}&  e^{-i\phi_2}&  e^{-i\phi_3}&  e^{-i\phi_4}& -e^{-i\phi_5}& -e^{-i\phi_6}& -e^{-i\phi_7}& -e^{-i\phi_8}
\end{matrix}\right)
\left(\begin{matrix}
\left|000\right\rangle\\ \left|001\right\rangle\\ \left|010\right\rangle\\ \left|011\right\rangle\\ \left|100\right\rangle\\ \left|101\right\rangle\\ \left|110\right\rangle\\ \left|111\right\rangle
\end{matrix}\right).
\end{eqnarray}
With these measurements basis, the quantum channel linking the three parties (i.e., equation (2)) can be written in the basis (147,258,369), as
\begin{eqnarray}
\left|\mathcal{F}\right\rangle
&=&\frac{1}{8}\left|\varrho_{147}\right\rangle_1\big[\left|\varsigma_{258}\right\rangle_1\left(\alpha_1e^{i\phi_1}\left|000\right\rangle+\alpha_2e^{i\phi_2}\left|001\right\rangle+\alpha_3e^{i\phi_3}\left|010\right\rangle+\alpha_4e^{\phi_4}\left|011\right\rangle
+\alpha_5e^{i\phi_5}\left|100\right\rangle+\alpha_6e^{i\phi_6}\right.\nonumber\\
&&\ \ \ \ \ \ \ \ \ \left.\times\left|101\right\rangle+\alpha_7e^{i\phi_7}\left|110\right\rangle+\alpha_8e^{i\phi_8}\left|111\right\rangle\right)+\left|\varsigma_{258}\right\rangle_2\left(\alpha_1e^{i\phi_1}\left|000\right\rangle-\alpha_2e^{i\phi_2}\left|001\right\rangle+\alpha_3e^{i\phi_3}\left|010\right\rangle\right.\nonumber\\
&&\ \ \ \ \ \ \ \ \ -\left.\alpha_4e^{i\phi_4}\left|011\right\rangle
+\alpha_5e^{i\phi_5}\left|100\right\rangle-\alpha_6e^{i\phi_6}\left|101\right\rangle
+\alpha_7e^{i\phi_7}\left|110\right\rangle-\alpha_8e^{i\phi_8}\left|111\right\rangle\right)+\left|\varsigma_{258}\right\rangle_3\left(\alpha_1e^{i\phi_1}\right.\nonumber\\
&&\ \ \ \ \ \ \ \ \ \left.\times\left|000\right\rangle-\alpha_2e^{i\phi_2}\left|001\right\rangle-\alpha_3e^{i\phi_3}\left|010\right\rangle+\alpha_4e^{i\phi_4}\left|011\right\rangle
-\alpha_5e^{i\phi_5}\left|100\right\rangle+\alpha_6e^{i\phi_6}\left|101\right\rangle
+\alpha_7e^{i\phi_7}\right.\nonumber\\
&&\ \ \ \ \ \ \ \ \ \left.\times\left|110\right\rangle-\alpha_8e^{i\phi_8}\left|111\right\rangle\right)+\left|\varsigma_{258}\right\rangle_4\left(\alpha_1e^{i\phi_1}\left|000\right\rangle+\alpha_2e^{i\phi_2}\left|001\right\rangle-\alpha_3e^{i\phi_3}\left|010\right\rangle-\alpha_4e^{i\phi_4}\left|011\right\rangle
\right.\nonumber\\
&&\ \ \ \ \ \ \ \ \ \left.+\alpha_5e^{i\phi_5}\left|100\right\rangle+\alpha_6e^{i\phi_6}\left|101\right\rangle
-\alpha_7e^{i\phi_7}\left|110\right\rangle-\alpha_8e^{i\phi_8}\left|111\right\rangle\right)+\left|\varsigma_{258}\right\rangle_5\left(\alpha_1e^{i\phi_1}\left|000\right\rangle-\alpha_2e^{i\phi_2}
\right.\nonumber\\
&&\ \ \ \ \ \ \ \ \ \left.\times\left|001\right\rangle+\alpha_3e^{i\phi_3}\left|010\right\rangle-\alpha_4e^{i\phi_4}\left|011\right\rangle-\alpha_5e^{i\phi_5}\left|100\right\rangle+\alpha_6e^{i\phi_6}\left|101\right\rangle-\alpha_7e^{i\phi_7}\left|110\right\rangle+\alpha_8e^{i\phi_8}\right.\nonumber\\
&&\ \ \ \ \ \ \ \ \ \left.\times\left|111\right\rangle\right)+\left|\varsigma_{258}\right\rangle_6\left(\alpha_1e^{i\phi_1}\left|000\right\rangle+\alpha_2e^{i\phi_2}\left|001\right\rangle-\alpha_3e^{i\phi_3}\left|010\right\rangle-\alpha_4e^{i\phi_4}\left|011\right\rangle-\alpha_5e^{i\phi_5}\left|100\right\rangle\right.\nonumber\\
&&\ \ \ \ \ \ \ \ \ \left.-\alpha_6e^{i\phi_6}\left|101\right\rangle
+\alpha_7e^{i\phi_7}\left|110\right\rangle+\alpha_8e^{i\phi_8}\left|111\right\rangle\right)+\left|\varsigma_{258}\right\rangle_7\left(\alpha_1e^{i\phi_1}\left|000\right\rangle-\alpha_2e^{i\phi_2}\left|001\right\rangle-\alpha_3e^{i\phi_3}\right.\nonumber\\
&&\ \ \ \ \ \ \ \ \ \left.\times\left|010\right\rangle+\alpha_4e^{i\phi_4}\left|011\right\rangle+\alpha_5e^{i\phi_5}\left|100\right\rangle-\alpha_6e^{i\phi_6}\left|101\right\rangle
-\alpha_7e^{i\phi_7}\left|110\right\rangle+\alpha_8e^{i\phi_8}\left|111\right\rangle\right)+\left|\varsigma_{258}\right\rangle_8\nonumber\\
&&\ \ \ \ \ \ \ \ \ \times\left(\alpha_1e^{i\phi_1}\left|000\right\rangle+\alpha_2e^{i\phi_2}\left|001\right\rangle+\alpha_3e^{i\phi_3}\left|010\right\rangle+\alpha_4e^{i\phi_4}\left|011\right\rangle
-\alpha_5e^{i\phi_5}\left|100\right\rangle-\alpha_6e^{i\phi_6}\left|101\right\rangle\right.\nonumber\\
&&\ \ \ \ \ \ \ \ \ \left.-\alpha_7e^{i\phi_7}\left|110\right\rangle-\alpha_8e^{i\phi_8}\left|111\right\rangle\right)\nonumber\\
&+&\frac{1}{8}\left|\varrho_{147}\right\rangle_2\big[\left|\varsigma_{258}\right\rangle_1\left(\alpha_2e^{i\phi_1}\left|000\right\rangle-\alpha_1e^{i\phi_2}\left|001\right\rangle+\alpha_4e^{i\phi_3}\left|010\right\rangle-\alpha_3e^{i\phi_4}\left|011\right\rangle
+\alpha_6e^{i\phi_5}\left|100\right\rangle-\alpha_5e^{i\phi_6}\right.\nonumber\\
&&\ \ \ \ \ \ \ \ \ \left.\times\left|101\right\rangle+\alpha_8e^{i\phi_7}\left|110\right\rangle-\alpha_7e^{i\phi_8}\left|111\right\rangle\right)+\left|\varsigma_{258}\right\rangle_2\left(\alpha_2e^{i\phi_1}\left|000\right\rangle+\alpha_1e^{i\phi_2}\left|001\right\rangle+\alpha_4e^{i\phi_3}\left|010\right\rangle\right.\nonumber\\
&&\ \ \ \ \ \ \ \ \ +\left.\alpha_3e^{i\phi_4}\left|011\right\rangle
+\alpha_6e^{i\phi_5}\left|100\right\rangle+\alpha_5e^{i\phi_6}\left|101\right\rangle
+\alpha_8e^{i\phi_7}\left|110\right\rangle+\alpha_7e^{i\phi_8}\left|111\right\rangle\right)+\left|\varsigma_{258}\right\rangle_3\left(\alpha_2e^{i\phi_1}\right.\nonumber\\
&&\ \ \ \ \ \ \ \ \ \left.\times\left|000\right\rangle+\alpha_1e^{i\phi_2}\left|001\right\rangle-\alpha_4e^{i\phi_3}\left|010\right\rangle-\alpha_3e^{i\phi_4}\left|011\right\rangle
-\alpha_6e^{i\phi_5}\left|100\right\rangle-\alpha_5e^{i\phi_6}\left|101\right\rangle
+\alpha_8e^{i\phi_7}\right.\nonumber\\
&&\ \ \ \ \ \ \ \ \ \left.+\left|110\right\rangle+\alpha_7e^{i\phi_8}\left|111\right\rangle\right)+\left|\varsigma_{258}\right\rangle_4\left(\alpha_2e^{i\phi_1}\left|000\right\rangle-\alpha_1e^{i\phi_2}\left|001\right\rangle-\alpha_4e^{i\phi_3}\left|010\right\rangle+\alpha_3e^{i\phi_4}\left|011\right\rangle
\right.\nonumber\\
&&\ \ \ \ \ \ \ \ \ \left.+\alpha_6e^{i\phi_5}\left|100\right\rangle-\alpha_5e^{i\phi_6}\left|101\right\rangle
-\alpha_8e^{i\phi_7}\left|110\right\rangle+\alpha_7e^{i\phi_8}\left|111\right\rangle\right)+\left|\varsigma_{258}\right\rangle_5\left(\alpha_2e^{i\phi_1}\left|000\right\rangle+\alpha_1e^{i\phi_2}
\right.\nonumber\\
&&\ \ \ \ \ \ \ \ \ \left.\times\left|001\right\rangle+\alpha_4e^{i\phi_3}\left|010\right\rangle+\alpha_3e^{i\phi_4}\left|011\right\rangle-\alpha_6e^{i\phi_5}\left|100\right\rangle-\alpha_5e^{i\phi_6}\left|101\right\rangle-\alpha_8e^{i\phi_7}\left|110\right\rangle-\alpha_7e^{i\phi_8}\right.\nonumber\\
&&\ \ \ \ \ \ \ \ \ \left.\times\left|111\right\rangle\right)+\left|\varsigma_{258}\right\rangle_6\left(\alpha_2e^{i\phi_1}\left|000\right\rangle-\alpha_1e^{i\phi_2}\left|001\right\rangle-\alpha_4e^{i\phi_3}\left|010\right\rangle+\alpha_3e^{i\phi_4}\left|011\right\rangle-\alpha_6e^{i\phi_5}\left|100\right\rangle\right.\nonumber\\
&&\ \ \ \ \ \ \ \ \ \left.+\alpha_5e^{i\phi_6}\left|101\right\rangle
+\alpha_8e^{i\phi_7}\left|110\right\rangle-\alpha_7e^{i\phi_8}\left|111\right\rangle\right)+\left|\varsigma_{258}\right\rangle_7\left(\alpha_2e^{i\phi_1}\left|000\right\rangle+\alpha_1e^{i\phi_2}\left|001\right\rangle-\alpha_4e^{i\phi_3}\right.\nonumber\\
&&\ \ \ \ \ \ \ \ \ \left.\times\left|010\right\rangle-\alpha_3e^{i\phi_4}\left|011\right\rangle-\alpha_6e^{i\phi_5}\left|100\right\rangle+\alpha_5e^{i\phi_6}\left|101\right\rangle
-\alpha_8e^{i\phi_7}\left|110\right\rangle+\alpha_7e^{i\phi_8}\left|111\right\rangle\right)+\left|\varsigma_{258}\right\rangle_8\nonumber\\
&&\ \ \ \ \ \ \ \ \ \times\left(\alpha_2e^{i\phi_1}\left|000\right\rangle-\alpha_1e^{i\phi_2}\left|001\right\rangle+\alpha_4e^{i\phi_3}\left|010\right\rangle-\alpha_3e^{i\phi_4}\left|011\right\rangle
-\alpha_6e^{i\phi_5}\left|100\right\rangle+\alpha_5e^{i\phi_6}\left|101\right\rangle\right.\nonumber\\
&&\ \ \ \ \ \ \ \ \ \left.-\alpha_8e^{i\phi_7}\left|110\right\rangle+\alpha_7e^{i\phi_8}\left|111\right\rangle\right)\nonumber\\
&+&\frac{1}{8}\left|\varrho_{147}\right\rangle_3\big[\left|\varsigma_{258}\right\rangle_1\left(\alpha_3e^{i\phi_1}\left|000\right\rangle-\alpha_4e^{i\phi_2}\left|001\right\rangle-\alpha_1e^{i\phi_3}\left|010\right\rangle+\alpha_2e^{i\phi_4}\left|011\right\rangle
-\alpha_7e^{i\phi_5}\left|100\right\rangle+\alpha_8e^{i\phi_6}\right.\nonumber\\
&&\ \ \ \ \ \ \ \ \ \left.\times\left|101\right\rangle+\alpha_5e^{i\phi_7}\left|110\right\rangle-\alpha_6e^{i\phi_8}\left|111\right\rangle\right)+\left|\varsigma_{258}\right\rangle_2\left(\alpha_3e^{i\phi_1}\left|000\right\rangle+\alpha_4e^{i\phi_2}\left|001\right\rangle-\alpha_1e^{i\phi_3}\left|010\right\rangle\right.\nonumber\\
&&\ \ \ \ \ \ \ \ \ -\left.\alpha_2e^{i\phi_4}\left|011\right\rangle
-\alpha_7e^{i\phi_5}\left|100\right\rangle-\alpha_8e^{i\phi_6}\left|101\right\rangle
+\alpha_5e^{i\phi_7}\left|110\right\rangle+\alpha_6e^{i\phi_8}\left|111\right\rangle\right)+\left|\varsigma_{258}\right\rangle_3\left(\alpha_3e^{i\phi_1}\right.\nonumber\\
&&\ \ \ \ \ \ \ \ \ \left.\times\left|000\right\rangle+\alpha_4e^{i\phi_2}\left|001\right\rangle+\alpha_1e^{i\phi_3}\left|010\right\rangle+\alpha_2e^{i\phi_4}\left|011\right\rangle
+\alpha_7e^{i\phi_5}\left|100\right\rangle+\alpha_8e^{i\phi_6}\left|101\right\rangle
+\alpha_5e^{i\phi_7}\right.\nonumber\\
&&\ \ \ \ \ \ \ \ \ \left.+\left|110\right\rangle+\alpha_6e^{i\phi_8}\left|111\right\rangle\right)+\left|\varsigma_{258}\right\rangle_4\left(\alpha_3e^{i\phi_1}\left|000\right\rangle-\alpha_4e^{i\phi_2}\left|001\right\rangle+\alpha_1e^{i\phi_3}\left|010\right\rangle-\alpha_2e^{i\phi_4}\left|011\right\rangle
\right.\nonumber\\
&&\ \ \ \ \ \ \ \ \ \left.-\alpha_7e^{i\phi_5}\left|100\right\rangle+\alpha_8e^{i\phi_6}\left|101\right\rangle
-\alpha_5e^{i\phi_7}\left|110\right\rangle+\alpha_6e^{i\phi_8}\left|111\right\rangle\right)+\left|\varsigma_{258}\right\rangle_5\left(\alpha_3e^{i\phi_1}\left|000\right\rangle+\alpha_4e^{i\phi_2}
\right.\nonumber\\
&&\ \ \ \ \ \ \ \ \ \left.\times\left|001\right\rangle-\alpha_1e^{i\phi_3}\left|010\right\rangle-\alpha_2e^{i\phi_4}\left|011\right\rangle+\alpha_7e^{i\phi_5}\left|100\right\rangle+\alpha_8e^{i\phi_6}\left|101\right\rangle-\alpha_5e^{i\phi_7}\left|110\right\rangle-\alpha_6e^{i\phi_8}\right.\nonumber\\
&&\ \ \ \ \ \ \ \ \ \left.\times\left|111\right\rangle\right)+\left|\varsigma_{258}\right\rangle_6\left(\alpha_3e^{i\phi_1}\left|000\right\rangle-\alpha_4e^{i\phi_2}\left|001\right\rangle+\alpha_1e^{i\phi_3}\left|010\right\rangle-\alpha_2e^{i\phi_4}\left|011\right\rangle+\alpha_7e^{i\phi_5}\left|100\right\rangle\right.\nonumber\\
&&\ \ \ \ \ \ \ \ \ \left.-\alpha_8e^{i\phi_6}\left|101\right\rangle
+\alpha_5e^{i\phi_7}\left|110\right\rangle-\alpha_6e^{i\phi_8}\left|111\right\rangle\right)+\left|\varsigma_{258}\right\rangle_7\left(\alpha_3e^{i\phi_1}\left|000\right\rangle+\alpha_4e^{i\phi_2}\left|001\right\rangle+\alpha_1e^{i\phi_3}\right.\nonumber\\
&&\ \ \ \ \ \ \ \ \ \left.\times\left|010\right\rangle+\alpha_2e^{i\phi_4}\left|011\right\rangle-\alpha_7e^{i\phi_5}\left|100\right\rangle-\alpha_8e^{i\phi_6}\left|101\right\rangle
-\alpha_5e^{i\phi_7}\left|110\right\rangle-\alpha_6e^{i\phi_8}\left|111\right\rangle\right)+\left|\varsigma_{258}\right\rangle_8\nonumber\\
&&\ \ \ \ \ \ \ \ \ \times\left(\alpha_3e^{i\phi_1}\left|000\right\rangle-\alpha_4e^{i\phi_2}\left|001\right\rangle-\alpha_1e^{i\phi_3}\left|010\right\rangle+\alpha_2e^{i\phi_4}\left|011\right\rangle
+\alpha_7e^{i\phi_5}\left|100\right\rangle-\alpha_8e^{i\phi_6}\left|101\right\rangle\right.\nonumber\\
&&\ \ \ \ \ \ \ \ \ \left.-\alpha_5e^{i\phi_7}\left|110\right\rangle+\alpha_6e^{i\phi_8}\left|111\right\rangle\right)\nonumber\\
&+&\frac{1}{8}\left|\varrho_{147}\right\rangle_4\big[\left|\varsigma_{258}\right\rangle_1\left(\alpha_4e^{i\phi_1}\left|000\right\rangle+\alpha_3e^{i\phi_2}\left|001\right\rangle-\alpha_2e^{i\phi_3}\left|010\right\rangle-\alpha_1e^{i\phi_4}\left|011\right\rangle
+\alpha_8e^{i\phi_5}\left|100\right\rangle+\alpha_7e^{i\phi_6}\right.\nonumber\\
&&\ \ \ \ \ \ \ \ \ \left.\times\left|101\right\rangle-\alpha_6e^{i\phi_7}\left|110\right\rangle-\alpha_5e^{i\phi_8}\left|111\right\rangle\right)+\left|\varsigma_{258}\right\rangle_2\left(\alpha_4e^{i\phi_1}\left|000\right\rangle-\alpha_3e^{i\phi_2}\left|001\right\rangle-\alpha_2e^{i\phi_3}\left|010\right\rangle\right.\nonumber\\
&&\ \ \ \ \ \ \ \ \ +\left.\alpha_1e^{i\phi_4}\left|011\right\rangle
+\alpha_8e^{i\phi_5}\left|100\right\rangle-\alpha_7e^{i\phi_6}\left|101\right\rangle
-\alpha_6e^{i\phi_7}\left|110\right\rangle+\alpha_5e^{i\phi_8}\left|111\right\rangle\right)+\left|\varsigma_{258}\right\rangle_3\left(\alpha_4e^{i\phi_1}\right.\nonumber\\
&&\ \ \ \ \ \ \ \ \ \left.\times\left|000\right\rangle-\alpha_3e^{i\phi_2}\left|001\right\rangle+\alpha_2e^{i\phi_3}\left|010\right\rangle-\alpha_1e^{i\phi_4}\left|011\right\rangle
-\alpha_8e^{i\phi_5}\left|100\right\rangle+\alpha_7e^{i\phi_6}\left|101\right\rangle
-\alpha_6e^{i\phi_7}\right.\nonumber\\
&&\ \ \ \ \ \ \ \ \ \left.+\left|110\right\rangle+\alpha_5e^{i\phi_8}\left|111\right\rangle\right)+\left|\varsigma_{258}\right\rangle_4\left(\alpha_4e^{i\phi_1}\left|000\right\rangle+\alpha_3e^{i\phi_2}\left|001\right\rangle+\alpha_2e^{i\phi_3}\left|010\right\rangle+\alpha_1e^{i\phi_4}\left|011\right\rangle
\right.\nonumber\\
&&\ \ \ \ \ \ \ \ \ \left.+\alpha_8e^{i\phi_5}\left|100\right\rangle+\alpha_7e^{i\phi_6}\left|101\right\rangle
+\alpha_6e^{i\phi_7}\left|110\right\rangle+\alpha_5e^{i\phi_8}\left|111\right\rangle\right)+\left|\varsigma_{258}\right\rangle_5\left(\alpha_4e^{i\phi_1}\left|000\right\rangle-\alpha_3e^{i\phi_2}
\right.\nonumber\\
&&\ \ \ \ \ \ \ \ \ \left.\times\left|001\right\rangle-\alpha_2e^{i\phi_3}\left|010\right\rangle+\alpha_1e^{i\phi_4}\left|011\right\rangle-\alpha_8e^{i\phi_5}\left|100\right\rangle+\alpha_7e^{i\phi_6}\left|101\right\rangle+\alpha_6e^{i\phi_7}\left|110\right\rangle-\alpha_5e^{i\phi_8}\right.\nonumber\\
&&\ \ \ \ \ \ \ \ \ \left.\times\left|111\right\rangle\right)+\left|\varsigma_{258}\right\rangle_6\left(\alpha_4e^{i\phi_1}\left|000\right\rangle+\alpha_3e^{i\phi_2}\left|001\right\rangle+\alpha_2e^{i\phi_3}\left|010\right\rangle+\alpha_1e^{i\phi_4}\left|011\right\rangle-\alpha_8e^{i\phi_5}\left|100\right\rangle\right.\nonumber\\
&&\ \ \ \ \ \ \ \ \ \left.-\alpha_7e^{i\phi_6}\left|101\right\rangle
-\alpha_6e^{i\phi_7}\left|110\right\rangle-\alpha_5e^{i\phi_8}\left|111\right\rangle\right)+\left|\varsigma_{258}\right\rangle_7\left(\alpha_4e^{i\phi_1}\left|000\right\rangle-\alpha_3e^{i\phi_2}\left|001\right\rangle+\alpha_2e^{i\phi_3}\right.\nonumber\\
&&\ \ \ \ \ \ \ \ \ \left.\times\left|010\right\rangle-\alpha_1e^{i\phi_4}\left|011\right\rangle+\alpha_8e^{i\phi_5}\left|100\right\rangle-\alpha_7e^{i\phi_6}\left|101\right\rangle
+\alpha_6e^{i\phi_7}\left|110\right\rangle-\alpha_5e^{i\phi_8}\left|111\right\rangle\right)+\left|\varsigma_{258}\right\rangle_8\nonumber\\
&&\ \ \ \ \ \ \ \ \ \times\left(\alpha_4e^{i\phi_1}\left|000\right\rangle+\alpha_3e^{i\phi_2}\left|001\right\rangle-\alpha_2e^{i\phi_3}\left|010\right\rangle-\alpha_1e^{i\phi_4}\left|011\right\rangle
-\alpha_8e^{i\phi_5}\left|100\right\rangle-\alpha_7e^{i\phi_6}\left|101\right\rangle\right.\nonumber\\
&&\ \ \ \ \ \ \ \ \ \left.+\alpha_6e^{i\phi_7}\left|110\right\rangle+\alpha_5e^{i\phi_8}\left|111\right\rangle\right)\nonumber\\
&+&\frac{1}{8}\left|\varrho_{147}\right\rangle_5\big[\left|\varsigma_{258}\right\rangle_1\left(\alpha_5e^{i\phi_1}\left|000\right\rangle-\alpha_6e^{i\phi_2}\left|001\right\rangle+\alpha_7e^{i\phi_3}\left|010\right\rangle-\alpha_8e^{i\phi_4}\left|011\right\rangle
-\alpha_1e^{i\phi_5}\left|100\right\rangle+\alpha_2e^{i\phi_6}\right.\nonumber\\
&&\ \ \ \ \ \ \ \ \ \left.\times\left|101\right\rangle-\alpha_3e^{i\phi_7}\left|110\right\rangle+\alpha_4e^{i\phi_8}\left|111\right\rangle\right)+\left|\varsigma_{258}\right\rangle_2\left(\alpha_5e^{i\phi_1}\left|000\right\rangle+\alpha_6e^{i\phi_2}\left|001\right\rangle+\alpha_7e^{i\phi_3}\left|010\right\rangle\right.\nonumber\\
&&\ \ \ \ \ \ \ \ \ +\left.\alpha_8e^{i\phi_4}\left|011\right\rangle
-\alpha_1e^{i\phi_5}\left|100\right\rangle-\alpha_2e^{i\phi_6}\left|101\right\rangle
-\alpha_3e^{i\phi_7}\left|110\right\rangle-\alpha_4e^{i\phi_8}\left|111\right\rangle\right)+\left|\varsigma_{258}\right\rangle_3\left(\alpha_5e^{i\phi_1}\right.\nonumber\\
&&\ \ \ \ \ \ \ \ \ \left.\times\left|000\right\rangle+\alpha_6e^{i\phi_2}\left|001\right\rangle-\alpha_7e^{i\phi_3}\left|010\right\rangle-\alpha_8e^{i\phi_4}\left|011\right\rangle
+\alpha_1e^{i\phi_5}\left|100\right\rangle+\alpha_2e^{i\phi_6}\left|101\right\rangle
-\alpha_3e^{i\phi_7}\right.\nonumber\\
&&\ \ \ \ \ \ \ \ \ \left.\times\left|110\right\rangle-\alpha_4e^{i\phi_8}\left|111\right\rangle\right)+\left|\varsigma_{258}\right\rangle_4\left(\alpha_5e^{i\phi_1}\left|000\right\rangle-\alpha_6e^{i\phi_2}\left|001\right\rangle-\alpha_7e^{i\phi_3}\left|010\right\rangle+\alpha_8e^{i\phi_4}\left|011\right\rangle
\right.\nonumber\\
&&\ \ \ \ \ \ \ \ \ \left.-\alpha_1e^{i\phi_5}\left|100\right\rangle+\alpha_2e^{i\phi_6}\left|101\right\rangle
+\alpha_3e^{i\phi_7}\left|110\right\rangle-\alpha_4e^{i\phi_8}\left|111\right\rangle\right)+\left|\varsigma_{258}\right\rangle_5\left(\alpha_5e^{i\phi_1}\left|000\right\rangle+\alpha_6e^{i\phi_2}
\right.\nonumber\\
&&\ \ \ \ \ \ \ \ \ \left.\times\left|001\right\rangle+\alpha_7e^{i\phi_3}\left|010\right\rangle+\alpha_8e^{i\phi_4}\left|011\right\rangle+\alpha_1e^{i\phi_5}\left|100\right\rangle+\alpha_2e^{i\phi_6}\left|101\right\rangle+\alpha_3e^{i\phi_7}\left|110\right\rangle+\alpha_4e^{i\phi_8}\right.\nonumber\\
&&\ \ \ \ \ \ \ \ \ \left.\times\left|111\right\rangle\right)+\left|\varsigma_{258}\right\rangle_6\left(\alpha_5e^{i\phi_1}\left|000\right\rangle-\alpha_6e^{i\phi_2}\left|001\right\rangle-\alpha_7e^{i\phi_3}\left|010\right\rangle+\alpha_8e^{i\phi_4}\left|011\right\rangle+\alpha_1e^{i\phi_5}\left|100\right\rangle\right.\nonumber\\
&&\ \ \ \ \ \ \ \ \ \left.-\alpha_2e^{i\phi_6}\left|101\right\rangle
-\alpha_3e^{i\phi_7}\left|110\right\rangle+\alpha_4e^{i\phi_8}\left|111\right\rangle\right)+\left|\varsigma_{258}\right\rangle_7\left(\alpha_5e^{i\phi_1}\left|000\right\rangle+\alpha_6e^{i\phi_2}\left|001\right\rangle-\alpha_7e^{i\phi_3}\right.\nonumber\\
&&\ \ \ \ \ \ \ \ \ \left.\times\left|010\right\rangle-\alpha_8e^{i\phi_4}\left|011\right\rangle-\alpha_1e^{i\phi_5}\left|100\right\rangle-\alpha_2e^{i\phi_6}\left|101\right\rangle
+\alpha_3e^{i\phi_7}\left|110\right\rangle+\alpha_4e^{i\phi_8}\left|111\right\rangle\right)+\left|\varsigma_{258}\right\rangle_8\nonumber\\
&&\ \ \ \ \ \ \ \ \ \times\left(\alpha_5e^{i\phi_1}\left|000\right\rangle-\alpha_6e^{i\phi_2}\left|001\right\rangle+\alpha_7e^{i\phi_3}\left|010\right\rangle-\alpha_8e^{i\phi_4}\left|011\right\rangle
+\alpha_1e^{i\phi_5}\left|100\right\rangle-\alpha_2e^{i\phi_6}\left|101\right\rangle\right.\nonumber\\
&&\ \ \ \ \ \ \ \ \ \left.+\alpha_3e^{i\phi_7}\left|110\right\rangle-\alpha_4e^{i\phi_8}\left|111\right\rangle\right)\nonumber\\
&+&\frac{1}{8}\left|\varrho_{147}\right\rangle_6\big[\left|\varsigma_{258}\right\rangle_1\left(\alpha_6e^{i\phi_1}\left|000\right\rangle+\alpha_5e^{i\phi_2}\left|001\right\rangle-\alpha_8e^{i\phi_3}\left|010\right\rangle-\alpha_7e^{i\phi_4}\left|011\right\rangle
-\alpha_2e^{i\phi_5}\left|100\right\rangle-\alpha_1e^{i\phi_6}\right.\nonumber\\
&&\ \ \ \ \ \ \ \ \ \left.\times\left|101\right\rangle+\alpha_4e^{i\phi_7}\left|110\right\rangle+\alpha_3e^{i\phi_8}\left|111\right\rangle\right)+\left|\varsigma_{258}\right\rangle_2\left(\alpha_6e^{i\phi_1}\left|000\right\rangle-\alpha_5e^{i\phi_2}\left|001\right\rangle-\alpha_8e^{i\phi_3}\left|010\right\rangle\right.\nonumber\\
&&\ \ \ \ \ \ \ \ \ +\left.\alpha_7e^{i\phi_4}\left|011\right\rangle
-\alpha_2e^{i\phi_5}\left|100\right\rangle+\alpha_1e^{i\phi_6}\left|101\right\rangle
+\alpha_4e^{i\phi_7}\left|110\right\rangle-\alpha_3e^{i\phi_8}\left|111\right\rangle\right)+\left|\varsigma_{258}\right\rangle_3\left(\alpha_6e^{i\phi_1}\right.\nonumber\\
&&\ \ \ \ \ \ \ \ \ \left.\times\left|000\right\rangle-\alpha_5e^{i\phi_2}\left|001\right\rangle+\alpha_8e^{i\phi_3}\left|010\right\rangle-\alpha_7e^{i\phi_4}\left|011\right\rangle
+\alpha_2e^{i\phi_5}\left|100\right\rangle-\alpha_1e^{i\phi_6}\left|101\right\rangle
+\alpha_4e^{i\phi_7}\right.\nonumber\\
&&\ \ \ \ \ \ \ \ \ \left.+\left|110\right\rangle-\alpha_3e^{i\phi_8}\left|111\right\rangle\right)+\left|\varsigma_{258}\right\rangle_4\left(\alpha_6e^{i\phi_1}\left|000\right\rangle+\alpha_5e^{i\phi_2}\left|001\right\rangle+\alpha_8e^{i\phi_3}\left|010\right\rangle+\alpha_7e^{i\phi_4}\left|011\right\rangle
-\right.\nonumber\\
&&\ \ \ \ \ \ \ \ \ \left.\alpha_2e^{i\phi_5}\left|100\right\rangle-\alpha_1e^{i\phi_6}\left|101\right\rangle
-\alpha_4e^{i\phi_7}\left|110\right\rangle-\alpha_3e^{i\phi_8}\left|111\right\rangle\right)+\left|\varsigma_{258}\right\rangle_5\left(\alpha_6e^{i\phi_1}\left|000\right\rangle-\alpha_5e^{i\phi_2}
+\right.\nonumber\\
&&\ \ \ \ \ \ \ \ \ \left.\times\left|001\right\rangle-\alpha_8e^{i\phi_3}\left|010\right\rangle+\alpha_7e^{i\phi_4}\left|011\right\rangle+\alpha_2e^{i\phi_5}\left|100\right\rangle+\alpha_6e^{i\phi_6}\left|101\right\rangle-\alpha_1e^{i\phi_7}\left|110\right\rangle-\alpha_4e^{i\phi_8}\right.\nonumber\\
&&\ \ \ \ \ \ \ \ \ \left.\times\left|111\right\rangle\right)+\left|\varsigma_{258}\right\rangle_6\left(\alpha_6e^{i\phi_1}\left|000\right\rangle+\alpha_5e^{i\phi_2}\left|001\right\rangle+\alpha_8e^{i\phi_3}\left|010\right\rangle+\alpha_7e^{i\phi_4}\left|011\right\rangle+\alpha_2e^{i\phi_5}\left|100\right\rangle\right.\nonumber\\
&&\ \ \ \ \ \ \ \ \ \left.+\alpha_1e^{i\phi_6}\left|101\right\rangle
+\alpha_4e^{i\phi_7}\left|110\right\rangle+\alpha_3e^{i\phi_8}\left|111\right\rangle\right)+\left|\varsigma_{258}\right\rangle_7\left(\alpha_6e^{i\phi_1}\left|000\right\rangle-\alpha_5e^{i\phi_2}\left|001\right\rangle+\alpha_8e^{i\phi_3}\right.\nonumber\\
&&\ \ \ \ \ \ \ \ \ \left.\times\left|010\right\rangle-\alpha_7e^{i\phi_4}\left|011\right\rangle-\alpha_2e^{i\phi_5}\left|100\right\rangle+\alpha_1e^{i\phi_6}\left|101\right\rangle
-\alpha_4e^{i\phi_7}\left|110\right\rangle+\alpha_3e^{i\phi_8}\left|111\right\rangle\right)+\left|\varsigma_{258}\right\rangle_8\nonumber\\
&&\ \ \ \ \ \ \ \ \ \times\left(\alpha_6e^{i\phi_1}\left|000\right\rangle+\alpha_5e^{i\phi_2}\left|001\right\rangle-\alpha_8e^{i\phi_3}\left|010\right\rangle-\alpha_7e^{i\phi_4}\left|011\right\rangle
+\alpha_2e^{i\phi_5}\left|100\right\rangle+\alpha_1e^{i\phi_6}\left|101\right\rangle\right.\nonumber\\
&&\ \ \ \ \ \ \ \ \ \left.-\alpha_4e^{i\phi_7}\left|110\right\rangle-\alpha_3e^{i\phi_8}\left|111\right\rangle\right)\nonumber\\
&+&\frac{1}{8}\left|\varrho_{147}\right\rangle_7\big[\left|\varsigma_{258}\right\rangle_1\left(\alpha_7e^{i\phi_1}\left|000\right\rangle-\alpha_8e^{i\phi_2}\left|001\right\rangle-\alpha_5e^{i\phi_3}\left|010\right\rangle+\alpha_6e^{i\phi_4}\left|011\right\rangle
+\alpha_3e^{i\phi_5}\left|100\right\rangle-\alpha_4e^{i\phi_6}\right.\nonumber\\
&&\ \ \ \ \ \ \ \ \ \left.\times\left|101\right\rangle-\alpha_1e^{i\phi_7}\left|110\right\rangle+\alpha_2e^{i\phi_8}\left|111\right\rangle\right)+\left|\varsigma_{258}\right\rangle_2\left(\alpha_7e^{i\phi_1}\left|000\right\rangle+\alpha_8e^{i\phi_2}\left|001\right\rangle-\alpha_5e^{i\phi_3}\left|010\right\rangle\right.\nonumber\\
&&\ \ \ \ \ \ \ \ \ -\left.\alpha_6e^{i\phi_4}\left|011\right\rangle
+\alpha_3e^{i\phi_5}\left|100\right\rangle+\alpha_4e^{i\phi_6}\left|101\right\rangle
-\alpha_1e^{i\phi_7}\left|110\right\rangle-\alpha_2e^{i\phi_8}\left|111\right\rangle\right)+\left|\varsigma_{258}\right\rangle_3\left(\alpha_7e^{i\phi_1}\right.\nonumber\\
&&\ \ \ \ \ \ \ \ \ \left.\times\left|000\right\rangle+\alpha_8e^{i\phi_2}\left|001\right\rangle+\alpha_5e^{i\phi_3}\left|010\right\rangle+\alpha_6e^{i\phi_4}\left|011\right\rangle
-\alpha_3e^{i\phi_5}\left|100\right\rangle-\alpha_4e^{i\phi_6}\left|101\right\rangle
-\alpha_1e^{i\phi_7}\right.\nonumber\\
&&\ \ \ \ \ \ \ \ \ \left.+\left|110\right\rangle-\alpha_2e^{i\phi_8}\left|111\right\rangle\right)+\left|\varsigma_{258}\right\rangle_4\left(\alpha_7e^{i\phi_1}\left|000\right\rangle-\alpha_8e^{i\phi_2}\left|001\right\rangle+\alpha_5e^{i\phi_3}\left|010\right\rangle-\alpha_6e^{i\phi_4}\left|011\right\rangle
\right.\nonumber\\
&&\ \ \ \ \ \ \ \ \ \left.+\alpha_3e^{i\phi_5}\left|100\right\rangle-\alpha_4e^{i\phi_6}\left|101\right\rangle
+\alpha_1e^{i\phi_7}\left|110\right\rangle-\alpha_2e^{i\phi_8}\left|111\right\rangle\right)+\left|\varsigma_{258}\right\rangle_5\left(\alpha_7e^{i\phi_1}\left|000\right\rangle+\alpha_8e^{i\phi_2}
\right.\nonumber\\
&&\ \ \ \ \ \ \ \ \ \left.\times\left|001\right\rangle-\alpha_5e^{i\phi_3}\left|010\right\rangle-\alpha_6e^{i\phi_4}\left|011\right\rangle-\alpha_3e^{i\phi_5}\left|100\right\rangle-\alpha_4e^{i\phi_6}\left|101\right\rangle+\alpha_1e^{i\phi_7}\left|110\right\rangle+\alpha_2e^{i\phi_8}\right.\nonumber\\
&&\ \ \ \ \ \ \ \ \ \left.\times\left|111\right\rangle\right)+\left|\varsigma_{258}\right\rangle_6\left(\alpha_7e^{i\phi_1}\left|000\right\rangle-\alpha_8e^{i\phi_2}\left|001\right\rangle+\alpha_5e^{i\phi_3}\left|010\right\rangle-\alpha_6e^{i\phi_4}\left|011\right\rangle-\alpha_3e^{i\phi_5}\left|100\right\rangle\right.\nonumber\\
&&\ \ \ \ \ \ \ \ \ \left.+\alpha_4e^{i\phi_6}\left|101\right\rangle
-\alpha_1e^{i\phi_7}\left|110\right\rangle+\alpha_2e^{i\phi_8}\left|111\right\rangle\right)+\left|\varsigma_{258}\right\rangle_7\left(\alpha_7e^{i\phi_1}\left|000\right\rangle+\alpha_8e^{i\phi_2}\left|001\right\rangle+\alpha_5e^{i\phi_3}\right.\nonumber\\
&&\ \ \ \ \ \ \ \ \ \left.\times\left|010\right\rangle+\alpha_6e^{i\phi_4}\left|011\right\rangle+\alpha_3e^{i\phi_5}\left|100\right\rangle+\alpha_4e^{i\phi_6}\left|101\right\rangle
+\alpha_1e^{i\phi_7}\left|110\right\rangle+\alpha_2e^{i\phi_8}\left|111\right\rangle\right)+\left|\varsigma_{258}\right\rangle_8\nonumber\\
&&\ \ \ \ \ \ \ \ \ \times\left(\alpha_7e^{i\phi_1}\left|000\right\rangle-\alpha_8e^{i\phi_2}\left|001\right\rangle-\alpha_5e^{i\phi_3}\left|010\right\rangle+\alpha_6e^{i\phi_4}\left|011\right\rangle
-\alpha_3e^{i\phi_5}\left|100\right\rangle+\alpha_4e^{i\phi_6}\left|101\right\rangle\right.\nonumber\\
&&\ \ \ \ \ \ \ \ \ \left.+\alpha_1e^{i\phi_7}\left|110\right\rangle-\alpha_2e^{i\phi_8}\left|111\right\rangle\right)\nonumber\\
&+&\frac{1}{8}\left|\varrho_{147}\right\rangle_8\big[\left|\varsigma_{258}\right\rangle_1\left(\alpha_8e^{i\phi_1}\left|000\right\rangle+\alpha_7e^{i\phi_2}\left|001\right\rangle+\alpha_6e^{i\phi_3}\left|010\right\rangle+\alpha_5e^{i\phi_4}\left|011\right\rangle
-\alpha_4e^{i\phi_5}\left|100\right\rangle-\alpha_3e^{i\phi_6}\right.\nonumber\\
&&\ \ \ \ \ \ \ \ \ \left.\times\left|101\right\rangle-\alpha_2e^{i\phi_7}\left|110\right\rangle-\alpha_1e^{i\phi_8}\left|111\right\rangle\right)+\left|\varsigma_{258}\right\rangle_2\left(\alpha_8e^{i\phi_1}\left|000\right\rangle-\alpha_7e^{i\phi_2}\left|001\right\rangle+\alpha_6e^{i\phi_3}\left|010\right\rangle\right.\nonumber\\
&&\ \ \ \ \ \ \ \ \ -\left.\alpha_5e^{i\phi_4}\left|011\right\rangle
-\alpha_4e^{i\phi_5}\left|100\right\rangle+\alpha_3e^{i\phi_6}\left|101\right\rangle
-\alpha_2e^{i\phi_7}\left|110\right\rangle+\alpha_1e^{i\phi_8}\left|111\right\rangle\right)+\left|\varsigma_{258}\right\rangle_3\left(\alpha_8e^{i\phi_1}\right.\nonumber\\
&&\ \ \ \ \ \ \ \ \ \left.\times\left|000\right\rangle-\alpha_7e^{i\phi_2}\left|001\right\rangle-\alpha_6e^{i\phi_3}\left|010\right\rangle+\alpha_5e^{i\phi_4}\left|011\right\rangle
+\alpha_4e^{i\phi_5}\left|100\right\rangle-\alpha_3e^{i\phi_6}\left|101\right\rangle
-\alpha_2e^{i\phi_7}\right.\nonumber\\
&&\ \ \ \ \ \ \ \ \ \left.+\left|110\right\rangle+\alpha_1e^{i\phi_8}\left|111\right\rangle\right)+\left|\varsigma_{258}\right\rangle_4\left(\alpha_8e^{i\phi_1}\left|000\right\rangle+\alpha_7e^{i\phi_2}\left|001\right\rangle-\alpha_6e^{i\phi_3}\left|010\right\rangle-\alpha_5e^{i\phi_4}\left|011\right\rangle
\right.\nonumber\\
&&\ \ \ \ \ \ \ \ \ \left.-\alpha_4e^{i\phi_5}\left|100\right\rangle-\alpha_3e^{i\phi_6}\left|101\right\rangle
+\alpha_2e^{i\phi_7}\left|110\right\rangle+\alpha_1e^{i\phi_8}\left|111\right\rangle\right)+\left|\varsigma_{258}\right\rangle_5\left(\alpha_8e^{i\phi_1}\left|000\right\rangle-\alpha_7e^{i\phi_2}
\right.\nonumber\\
&&\ \ \ \ \ \ \ \ \ \left.\times\left|001\right\rangle+\alpha_6e^{i\phi_3}\left|010\right\rangle-\alpha_5e^{i\phi_4}\left|011\right\rangle+\alpha_4e^{i\phi_5}\left|100\right\rangle-\alpha_3e^{i\phi_6}\left|101\right\rangle+\alpha_2e^{i\phi_7}\left|110\right\rangle-\alpha_1e^{i\phi_8}\right.\nonumber\\
&&\ \ \ \ \ \ \ \ \ \left.\times\left|111\right\rangle\right)+\left|\varsigma_{258}\right\rangle_6\left(\alpha_8e^{i\phi_1}\left|000\right\rangle+\alpha_7e^{i\phi_2}\left|001\right\rangle-\alpha_6e^{i\phi_3}\left|010\right\rangle-\alpha_5e^{i\phi_4}\left|011\right\rangle+\alpha_4e^{i\phi_5}\left|100\right\rangle\right.\nonumber\\
&&\ \ \ \ \ \ \ \ \ \left.+\alpha_3e^{i\phi_6}\left|101\right\rangle
-\alpha_2e^{i\phi_7}\left|110\right\rangle-\alpha_1e^{i\phi_8}\left|111\right\rangle\right)+\left|\varsigma_{258}\right\rangle_7\left(\alpha_8e^{i\phi_1}\left|000\right\rangle-\alpha_7e^{i\phi_2}\left|001\right\rangle-\alpha_6e^{i\phi_3}\right.\nonumber\\
&&\ \ \ \ \ \ \ \ \ \left.\times\left|010\right\rangle+\alpha_5e^{i\phi_4}\left|011\right\rangle-\alpha_4e^{i\phi_5}\left|100\right\rangle+\alpha_3e^{i\phi_6}\left|101\right\rangle
+\alpha_2e^{i\phi_7}\left|110\right\rangle-\alpha_1e^{i\phi_8}\left|111\right\rangle\right)+\left|\varsigma_{258}\right\rangle_8\nonumber\\
&&\ \ \ \ \ \ \ \ \ \times\left(\alpha_8e^{i\phi_1}\left|000\right\rangle+\alpha_7e^{i\phi_2}\left|001\right\rangle+\alpha_6e^{i\phi_3}\left|010\right\rangle+\alpha_5e^{i\phi_4}\left|011\right\rangle
+\alpha_4e^{i\phi_5}\left|100\right\rangle+\alpha_3e^{i\phi_6}\left|101\right\rangle\right.\nonumber\\
&&\ \ \ \ \ \ \ \ \ \left.+\alpha_2e^{i\phi_7}\left|110\right\rangle+\alpha_1e^{i\phi_8}\left|111\right\rangle\right).\nonumber\\ \label{E5}
\end{eqnarray}
MATLAB can be use to substantiate flawless of the above equation (\ref{E5}). Now, after the measurements have been completed, Alice and Bob send information about their measurements to receiver Chika via classical channel. As it can be clearly seen in equation (\ref{E5}), if Alice's measurement is $\left|\varrho_{147}\right\rangle_1$ and Bob's measurements are $\{\left|\varsigma_{258}\right\rangle_n, n=1-8 \}$, then by effectuating a unitary transformation on particles trio $(3,6,9)$, Chika can reconstruct the state of the particle which Alice and Bob intend to prepare for her. For instance, let us consider that Bob's projective measurement is $\left|\varsigma_{258}\right\rangle_6$, then the state of particles trio (3,6,9) will collapse to $\alpha_1e^{-i\phi_1}\left|000\right\rangle+\alpha_2e^{-i\phi_2}\left|001\right\rangle-
 \alpha_3e^{-i\phi_3}\left|010\right\rangle-\alpha_4e^{-i\phi_4}\left|011\right\rangle-
 \alpha_5e^{-i\phi_5}\left|100\right\rangle-\alpha_6e^{-i\phi_6}\left|101\right\rangle+ 
 \alpha_7e^{-i\phi_7}\left|110\right\rangle+\alpha_8e^{-i\phi_8}\left|111\right\rangle$. Applying unitary operation $\sigma_z\otimes\sigma_z\otimes I$, Chika can reconstruct the state $\Omega$.  However, regarding the measurement of other states (i.e., Alice's outcome measurements  $\{\left|\varrho_{147}\right\rangle_r, r=2-8\}$ and Bob's outcome measurements $\{\left|\varsigma_{258}\right\rangle_n, n=1-8 \}$ ), the JRSP fails. Thus, the probability of success is $12.5\%$. However, this probability of achieving success can be ameliorated by considering some special cases as shown in Table 1.

\begin{table}[h!]
{\scriptsize
\caption{\footnotesize Special cases to ameliorate the success probability, state obtain after changing of variables and the appropriate unitary transformation that will be utilized by Chika in order to reconstruct the state of the particle which Alice and Bob intend to prepare for her from the shared entangled state. For a particular $\left|\varrho_{147}\right\rangle_r, r=2-8$, we only consider $\left|\varsigma_{258}\right\rangle_1$. However, for other states namely $\left|\varsigma_{258}\right\rangle_t, t\in (2-8)$, similar approach can be employed.} \vspace*{10pt}{\footnotesize
\begin{tabular}{M{1cm}|M{2.8cm}|M{10cm}|M{2.5cm}|}\hline\hline
{}&{}&{}&{}\\[-1.0ex]
Case& Variable transfor -mation& State obtained after variable transformation& Unitary transf- ormation ($UT$) \\[3.5ex]\hline
$r=2$&$\alpha_1 = \alpha_2$, $\alpha_3 = \alpha_4$, \newline $\alpha_5 = \alpha_6$, $\alpha_6 = \alpha_7.$&
$\alpha_1e^{i\phi_1}\left|000\right\rangle-\alpha_2e^{i\phi_2}\left|001\right\rangle+
 \alpha_3e^{i\phi_3}\left|010\right\rangle-\alpha_4e^{i\phi_4}\left|011\right\rangle+ \newline
 \alpha_5e^{i\phi_5}\left|100\right\rangle-\alpha_6e^{i\phi_6}\left|101\right\rangle+ 
 \alpha_7e^{i\phi_7}\left|110\right\rangle-\alpha_8e^{i\phi_8}\left|111\right\rangle$
 &$I\otimes I\otimes\sigma_z$\\[6.5ex]\hline
$r=3$&$\alpha_1 = \alpha_3$, $\alpha_2 = \alpha_4$, \newline $\alpha_5 = \alpha_7$, $\alpha_6 = \alpha_8.$&
$\alpha_1e^{i\phi_1}\left|000\right\rangle-\alpha_2e^{i\phi_2}\left|001\right\rangle-
 \alpha_3e^{i\phi_3}\left|010\right\rangle+\alpha_4e^{i\phi_4}\left|011\right\rangle- \newline
 \alpha_5e^{i\phi_5}\left|100\right\rangle+\alpha_6e^{i\phi_6}\left|101\right\rangle+ 
 \alpha_7e^{i\phi_7}\left|110\right\rangle-\alpha_8e^{i\phi_8}\left|111\right\rangle$
 &$\sigma_z\otimes\sigma_z\otimes\sigma_z$\\[6.5ex]\hline
$r=4$&$\alpha_1 = \alpha_4$, $\alpha_2 = \alpha_3$, \newline $\alpha_5 = \alpha_8$, $\alpha_6 = \alpha_7.$&
$\alpha_1e^{i\phi_1}\left|000\right\rangle+\alpha_2e^{i\phi_2}\left|001\right\rangle-
 \alpha_3e^{i\phi_3}\left|010\right\rangle-\alpha_4e^{i\phi_4}\left|011\right\rangle+ \newline
 \alpha_5e^{i\phi_5}\left|100\right\rangle+\alpha_6e^{i\phi_6}\left|101\right\rangle-
 \alpha_7e^{i\phi_7}\left|110\right\rangle-\alpha_8e^{i\phi_8}\left|111\right\rangle$
 &$I\otimes\sigma_z\otimes I$\\[6.5ex]\hline
$r=5$&$\alpha_1 = \alpha_5$, $\alpha_2 = \alpha_6$, \newline $\alpha_3 = \alpha_7$, $\alpha_4 = \alpha_8.$&
$\alpha_1e^{i\phi_1}\left|000\right\rangle-\alpha_2e^{i\phi_2}\left|001\right\rangle+
 \alpha_3e^{i\phi_3}\left|010\right\rangle-\alpha_4e^{i\phi_4}\left|011\right\rangle- \newline
 \alpha_5e^{i\phi_5}\left|100\right\rangle+\alpha_6e^{i\phi_6}\left|101\right\rangle- 
 \alpha_7e^{i\phi_7}\left|110\right\rangle+\alpha_8e^{i\phi_8}\left|111\right\rangle$
 &$\sigma_z\otimes I\otimes\sigma_z$\\[6.5ex]\hline
$r=6$&$\alpha_1 = \alpha_6$, $\alpha_2 = \alpha_5$, \newline $\alpha_3 = \alpha_8$, $\alpha_4 = \alpha_7.$&
$\alpha_1e^{i\phi_1}\left|000\right\rangle+\alpha_2e^{i\phi_2}\left|001\right\rangle-
 \alpha_3e^{i\phi_3}\left|010\right\rangle-\alpha_4e^{i\phi_4}\left|011\right\rangle- \newline
 \alpha_5e^{i\phi_5}\left|100\right\rangle-\alpha_6e^{i\phi_6}\left|101\right\rangle+ 
 \alpha_7e^{i\phi_7}\left|110\right\rangle+\alpha_8e^{i\phi_8}\left|111\right\rangle$
 &$\sigma_z\otimes\sigma_z\otimes I$\\[6.5ex]\hline
$r=7$&$\alpha_1 = \alpha_7$, $\alpha_2 = \alpha_8$, \newline $\alpha_5 = \alpha_3$, $\alpha_4 = \alpha_6.$&
$\alpha_1e^{i\phi_1}\left|000\right\rangle-\alpha_2e^{i\phi_2}\left|001\right\rangle-
 \alpha_3e^{i\phi_3}\left|010\right\rangle+\alpha_4e^{i\phi_4}\left|011\right\rangle+ \newline
 \alpha_5e^{i\phi_5}\left|100\right\rangle-\alpha_6e^{i\phi_6}\left|101\right\rangle- 
 \alpha_7e^{i\phi_7}\left|110\right\rangle+\alpha_8e^{i\phi_8}\left|111\right\rangle$
 &$I\otimes\sigma_z\otimes\sigma_z$\\[6.5ex]\hline
$r=8$&$\alpha_1 = \alpha_8$, $\alpha_2 = \alpha_7$, \newline $\alpha_3 = \alpha_6$, $\alpha_4 = \alpha_5.$&
$\alpha_1e^{i\phi_1}\left|000\right\rangle+\alpha_2e^{i\phi_2}\left|001\right\rangle+
 \alpha_3e^{i\phi_3}\left|010\right\rangle+\alpha_4e^{i\phi_4}\left|011\right\rangle- \newline
 \alpha_5e^{i\phi_5}\left|100\right\rangle-\alpha_6e^{i\phi_6}\left|101\right\rangle-
 \alpha_7e^{i\phi_7}\left|110\right\rangle-\alpha_8e^{i\phi_8}\left|111\right\rangle$&$\sigma_z\otimes I\otimes I$\\[6.5ex]\hline
 \end{tabular}\label{tab1}}}
\end{table}

Each cases considered in Table \ref{tab1} will increase the success probability of this JRSP scheme to $25\%$. Moreover, suppose in equation (\ref{E5}), $\alpha_1=\alpha_2=...=\alpha_8=\frac{1}{2\sqrt{2}}$ or $\phi_1=\phi_2=...=\phi_8=0$, then, Chika can easily apply an appropriate unitary transformation on her particle trio $(3,6,9)$ to realize the original state $\left|\Omega\right\rangle$. In this case, the success probability for the JRSP will be $100\%$. In the next section, we study the influence of noises on this protocol. This have also been delineated in Figure 1.

\section{JRSP of an arbitrary three-particle state in noisy environment}
The unwanted interactions of the surroundings with quantum communication constitute what we referred to as quantum noise. Quantum systems losses their properties as a consequence of these interactions. The effects of quantum noises on RSP have been investigated experimentally in reference \cite{A10} and theoretically, the effects of quantum noises on RSP and JRSP have been reported in references \cite{A19,A20}. It is worth mentioning that these efforts were two-qubit based. To the best of our knowledge, no report so far on JRSP of an arbitrary three-qubit in noisy environment. It is therefore the priority purpose of this section to examine this. We shall consider amplitude-damping, phase-damping and depolarizing quantum noisy channels as three models for the noise. We derive the analytical expression for fidelity in each cases. 
\begin{figure*}[!t]
\centering \includegraphics[height=40mm, width=110mm]{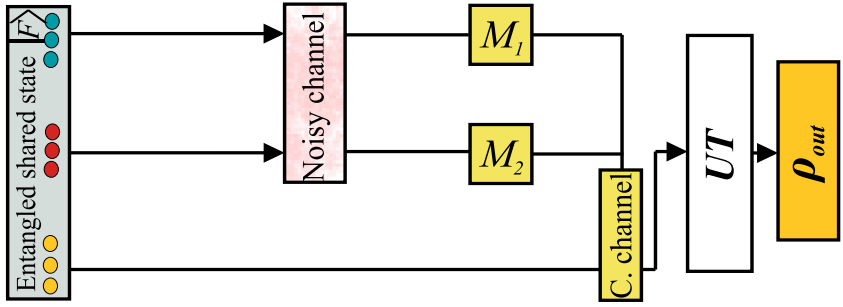}
\caption{\protect\small The top wire represents Alice's system, the middle wire denotes Bob's system and the bottom wire is for Chika's system. Alice's and Bob's qubits are distributed in quantum noisy channel. Alice's measurement is denoted as $M_1=\left|\varrho_{147}\right\rangle_1\ {}_1\left\langle\varrho_{147}\right|$ while Bob's measurement is denoted as  $M_2=\left|\varsigma_{258}\right\rangle_n\ {}_n\left\langle\varsigma_{258}\right|$. ``C. Channel" is the channel utilized by Alice and Bob for classical communication with the remote receiver Chika. Using the information received from Alice and Bob, Chika can apply $UT$ to retrieve $\left|\Omega\right\rangle$.} 
\label{fig1}
\end{figure*}
\subsection{JRSP in amplitude-damping environment}
In this subsection, we examine the effect of amplitude-damping noise on the JRSP of an arbitrary three-particle state. Amplitude-damping noise represents one of the valuable decoherence noise which provides us with description of energy-dissipation effects. For instance, an
atom which spontaneously emits a photon. The general behavior of this noise is characterized by the following set of Kraus operators \cite{A21,A22}
\begin{equation}
E_0^A=\left[\begin{matrix}1&0\\0&\sqrt{1-\eta_{_A}}\end{matrix}\right], \ \ \ \mbox{and}\ \ \ \ E_1^A=\left[\begin{matrix}0&\sqrt{\eta_{_A}}\\0&0\end{matrix}\right],
\end{equation}
where $\eta_{_A}(0\leq\eta_{_A}\leq 1)$ denotes the decoherence rate which describes the probability error of amplitude-damping when a particle passes through a noisy environment. Now, considering the fact that qubits trio $(3,6,9)$ are not transmitted through noisy channel, we express the effect of the amplitude damping noise on the shared entangled state as
\begin{equation}
\mathcal{A}(\rho)=\sum_{i,j}E_i^{A^1}\otimes E_i^{A^4}\otimes E_i^{A^7}\otimes E_j^{A^2}\otimes E_j^{A^5}\otimes E_j^{A^8}\rho\left(E_i^{A^1}\otimes E_i^{A^4}\otimes E_i^{A^7}\otimes E_j^{A^2}\otimes E_j^{A^5}\otimes E_j^{A^8}\right)^\dag,
\end{equation}
where $\rho=\left|\mathcal{F}\right\rangle\left\langle\mathcal{F}\right|$, $i,j\in\{0,1\}$ and the superscripts $(^{147258})$ denote the action of operator $E$ on which qubit.  From the previous section, we showed that JRSP will only be successful if Alice's measurement is $\left|\varrho_{147}\right\rangle_1$ and Bob's measurements are $\{\left|\varsigma_{258}\right\rangle_n, n=1-8 \}$ except if we consider some special cases. This is also appurtenant under noisy condition. It then insinuates that the failure cases cannot be considered as output state  and ergo; the shared state becomes a mixed state after particles distribution. Thus we express this as
\begin{eqnarray}
\mathcal{A}(\rho)_{123456789}&=&\frac{1}{8}\big[ \left|000000000\right\rangle+(1-\eta_{_A})\left|000111000\right\rangle+(1-\eta_{_A})\left|111000000\right\rangle+(1-\eta_{_A})^2\nonumber\\
&&\ \times\left|111111000\right\rangle+(1-\eta_{_A})\left|000000111\right\rangle+(1-\eta_{_A})^2\left|000111111\right\rangle+(1-\eta_{_A})^2\nonumber\\
&&\
\times\left|111000111\right\rangle+(1-\eta_{_A})^3\left|111111111\right\rangle\big]\times\big[\left\langle 000000000\right|+(1-\eta_{_A})\left\langle 000111000\right|\nonumber\\
&&\ 
+(1-\eta_{_A})\left\langle111000000\right|+(1-\eta_{_A})^2\left\langle 111111000\right|+(1-\eta_{_A})\left\langle 000000111\right|+(1-\eta_{_A})^2\nonumber\\
&&\
\times\left\langle000111111\right|+(1-\eta_{_A})^2\left\langle111000111\right|+(1-\eta_{_A})^3\left\langle 111111111\right|\big]+(1-\eta_{_A})^{3}\eta_{_A}^{3}
\nonumber\\
&&\
\times\left|101101101\right\rangle\left\langle101101101\right|+(1-\eta_{_A})^{3}\eta_{_A}^{3}\left|011011011\right\rangle\left\langle011011011\right|\nonumber\\
&&\
+\eta_{_A}^6\left|001001001\right\rangle\left\langle001001001\right|,
\end{eqnarray}
and the density matrix of the final state becomes
\begin{equation}
\rho_{out}=Tr_{147,258}\{U_0\mathcal{A}(\rho)U_0^\dag\},
\label{E16}
\end{equation}
where $U_0\left[=\left(\left|\varrho\right\rangle_{147}\ _{147}\left\langle\varrho\right|\otimes I_{258}\otimes I_{369}\right)\left(I_{147}\otimes \left|\varsigma\right\rangle_{258}\ _{258}\left\langle\varsigma\right|\otimes I_{369}\right)\left(I_{147}\otimes I_{258}\otimes UT^n_{369}\right),\right]$ denotes the unitary operator to complete the JRSP process and $Tr_{147,258}$ represents partial trace over particles trios $(1,4,7)$ and $(2,5,8)$. Thus, using equation (\ref{E16}), the density matrix of the output state (in the basis $369$) becomes
\begin{eqnarray}
\rho_{out}&=&\bigg[ \alpha_1e^{i\phi_1}\left|000\right\rangle+\alpha_2e^{i\phi_2}(1-\eta_{_A})\left|001\right\rangle+\alpha_3e^{i\phi_3}(1-\eta_{_A})\left|010\right\rangle+\alpha_4e^{i\phi_4}(1-\eta_{_A})^2\left|011\right\rangle+\alpha_5e^{i\phi_5}\nonumber\\
&& \times(1-\eta_{_A})\left|100\right\rangle+\alpha_6e^{i\phi_6}(1-\eta_{_A})^2\left|101\right\rangle+\alpha_7e^{i\phi_7}(1-\eta_{_A})^2\left|110\right\rangle+\alpha_8e^{i\phi_8}(1-\eta_{_A})^3\left|111\right\rangle\bigg]\nonumber\\
&&
\times\bigg[\alpha_1e^{-i\phi_1}\left\langle000\right|+\alpha_2e^{-i\phi_2}(1-\eta_{_A})\left\langle001\right|+\alpha_3e^{-i\phi_3}(1-\eta_{_A})\left\langle010\right|+\alpha_4e^{-i\phi_4}(1-\eta_{_A})^2\left\langle011\right|\nonumber\\
&& +\alpha_5e^{-i\phi_5}(1-\eta_{_A})\left\langle100\right|+\alpha_6e^{-i\phi_6}(1-\eta_{_A})^2\left\langle101\right|+\alpha_7e^{-i\phi_7}(1-\eta_{_A})^2\left\langle110\right|+\alpha_8e^{-i\phi_8}(1-\eta_{_A})^3\nonumber\\
&& \times\left\langle111\right|\bigg]+\alpha_8^2(1-\eta_{_A})^3\eta_{_A}^3\left|111\right\rangle\left\langle111\right|+\alpha_1^2(1-\eta_{_A})^3\eta_{_A}^3\left|111\right\rangle\left\langle111\right|+\alpha_1^2\eta_{_A}^6\left|111\right\rangle\left\langle111\right|.
\end{eqnarray}
Now, in order to determine closeness of the final state to the initial state, we use the fidelity $F=\left\langle\Omega\right|\rho_{out}\left|\Omega\right\rangle$ \cite{A20}. Thus, we have
\begin{eqnarray}
F=\bigg[\alpha_1^2+(1-\eta_{_A})\left(\alpha_2^2+\alpha_3^2+\alpha_5^2\right)+(1-\eta_{_A})^2\left(\alpha_4^2+\alpha_6^2+\alpha_7^2\right)+(1-\eta_{_A})^3\alpha_8^2\bigg]^2\nonumber\\
+\alpha_8^4(1-\eta_{_A})^3\eta_{_A}^3+\alpha_8^2\alpha_1^2(1-\eta_{_A})^3\eta_{_A}^3+\alpha_1^2\alpha_8^2\eta_{_A}^6,
\end{eqnarray}
which is untrammeled of phase parameter but depend on amplitude factor and the decoherence rate. For $\alpha_1=\alpha_2=...=\alpha_8=1/\left(2\sqrt{2}\right)$, and $\eta_{_A}=0$, then $F=1$ which is perfect JRSP while for $\eta_{_A}=1$, $F=1/32$. Figure 2 (a) shows 3D-plot of fidelity as a function of $\alpha_1$ and $\eta_{_A}$. We observe that the fidelity dwindle as $\eta_{_A}$ increase.
\subsection{JRSP in phase-damping channel}
In this subsection, we scrutinize the effect of phase-damping noise on the JRSP of an arbitrary three-particle state. Phase-damping noise describes the loss of quantum information without loss of energy. Phase-damping noise provides a revealing caricature of decoherence in realistic physical situations, with all inessential mathematical details stripped away. An example of this is randomly scattering of photon as it transverse through a waveguide \cite{A23}. The energy eigenstate does not vary as a function of time, instead it amasses phase which commensurates with eigenvalue. Consequently, the limited information regarding the relative phase between energy eigenstates is lost when the evolution time is not known. The behavior of this noise is characterized by the following set of Kraus operators \cite{A21,A22}
\begin{equation}
E_0^P=\sqrt{1-\eta_{_P}}{\bf I},\ \ \ \ E_1^P=\sqrt{\eta_{_P}}\left[\begin{matrix}1&0\\0&0\end{matrix}\right]\ \ \ \mbox{and}\ \ \ \ E_2^P=\sqrt{\eta_{_P}}\left[\begin{matrix}0&0\\0&1\end{matrix}\right],
\end{equation}
where $\eta_{_P}(0\leq\eta_{_P}\leq 1)$ denotes the decoherence rate for the phase-damping noise. Now, considering the fact that qubits trio $(3,6,9)$ are not transmitted through noisy channel, we express the effect of the phase-damping noise on the shared entangled state as
\begin{equation}
\mathcal{P}(\rho)=\sum_{i,j}E_i^{P^1}\otimes E_i^{P^4}\otimes E_i^{P^7}\otimes E_j^{P^2}\otimes E_j^{P^5}\otimes E_j^{P^8}\rho\left(E_i^{P^1}\otimes E_i^{P^4}\otimes E_i^{P^7}\otimes E_j^{P^2}\otimes E_j^{P^5}\otimes E_j^{P^8}\right)^\dag,
\end{equation}
where $\rho=\left|\mathcal{F}\right\rangle\left\langle\mathcal{F}\right|$, $i,j\in\{0,1,2\}$ and the superscripts $(^{147258})$ denote the action of operator $E$ on which qubit. The shared state after particles distribution becomes
\begin{eqnarray}
\mathcal{P}(\rho)_{123456789}&=&\frac{(1-\eta_{_P})^6}{8}\big[
 \left|000000000\right\rangle+\left|000111000\right\rangle+\left|111000000\right\rangle+\left|111111000\right\rangle+\left|000000111\right\rangle\nonumber\\
&&\ \ \ \ \ \ \ \ \ \ \ \ \
+\left|000111111\right\rangle+\left|111000111\right\rangle+\left|111111111\right\rangle\big]\times\big[
\left\langle000000000\right|+\left\langle000111000\right|\nonumber\\
&&\ \ \ \ \ \ \ \ \ \ \ \ \
+\left\langle111000000\right|+\left\langle111111000\right|+\left\langle000000111\right|+\left\langle000111111\right|+\left\langle111000111\right|\nonumber\\
&&\ \ \ \ \ \ \ \ \ \ \ \ \
+\left\langle111111111\right|\big]+\big[2\eta_{_P}^{3}(1-\eta_{_P})^{3}+\eta_{_P}^6\big]\big[\left|000000000\right\rangle\left\langle000000000\right|\nonumber\\
&&\ \ \ \ \ \ \ \ \ \ \ \
+\left|111111111\right\rangle\left\langle111111111\right|\big],
\end{eqnarray}
and the density matrix of the final state can be calculated from (\ref{E16}) as
\begin{eqnarray}
\rho_{out}&=&(1-\eta_{_P})^6\bigg[ \alpha_1e^{i\phi_1}\left|000\right\rangle+\alpha_2e^{i\phi_2}\left|001\right\rangle+\alpha_3e^{i\phi_3}\left|010\right\rangle+\alpha_4e^{i\phi_4}\left|011\right\rangle+\alpha_5e^{i\phi_5}\left|100\right\rangle+\alpha_6e^{i\phi_6}\left|101\right\rangle\nonumber\\
&&\ \ \ \ \ \ \ \ \ \ \ \ +\alpha_7e^{i\phi_7}\left|110\right\rangle+\alpha_8e^{i\phi_8}\left|111\right\rangle\bigg]\times\bigg[ \alpha_1e^{-i\phi_1}\left\langle000\right|+\alpha_2e^{-i\phi_2}\left\langle001\right|+\alpha_3e^{-i\phi_3}\left\langle 010\right|+\alpha_4e^{-i\phi_4}\nonumber\\
&&\ \ \ \ \ \ \ \ \ \ \ \ \times\left\langle 011\right|+\alpha_5e^{-i\phi_5}\left\langle 100\right|+\alpha_6e^{-i\phi_6}\left\langle101\right|+\alpha_7e^{-i\phi_7}\left\langle 110\right|+\alpha_8e^{-i\phi_8}\left\langle 111\right|\bigg]\nonumber\\
&&\ \ \ \ \ \ \ \ \ \ \ \ +\bigg[2\eta_{_P}^3(1-\eta_{_P})^3+\eta_{_P}^6\bigg]\times\bigg[\alpha_1^2\left|000\right\rangle\left\langle000\right|+\alpha_8^2\left|111\right\rangle\left\langle111\right|\bigg].
\end{eqnarray}
The closeness of the final state to the initial state can now be determined using the fidelity $F=\left\langle\Omega\right|\rho_{out}\left|\Omega\right\rangle$ \cite{A20}. Thus, we have
\begin{eqnarray}
F=(1-\eta_{_P})^6+\left(\alpha_1^4+\alpha_8^4\right)\bigg[2\eta_{_P}^3(1-\eta_{_P})^3+\eta_{_P}^6\bigg],
\end{eqnarray}
which is also independent of phase parameter but depends on the amplitude factor and the decoherence rate. For $\alpha_1=\alpha_2=...=\alpha_8=1/\left(2\sqrt{2}\right)$, and $\eta_{_P}=0$, then $F=1$ which symbolize perfect JRSP while for $\eta_{_P}=1$, $F=1/32$. Figure 2 (b) shows the plot of fidelity as a function of $\alpha_1$ and $\eta_{_P}$. As it can be seen, the fidelity also diminishes as $\eta_{_P}$ increases.
\begin{figure*}[!t]
\centering \includegraphics[height=60mm, width=160mm]{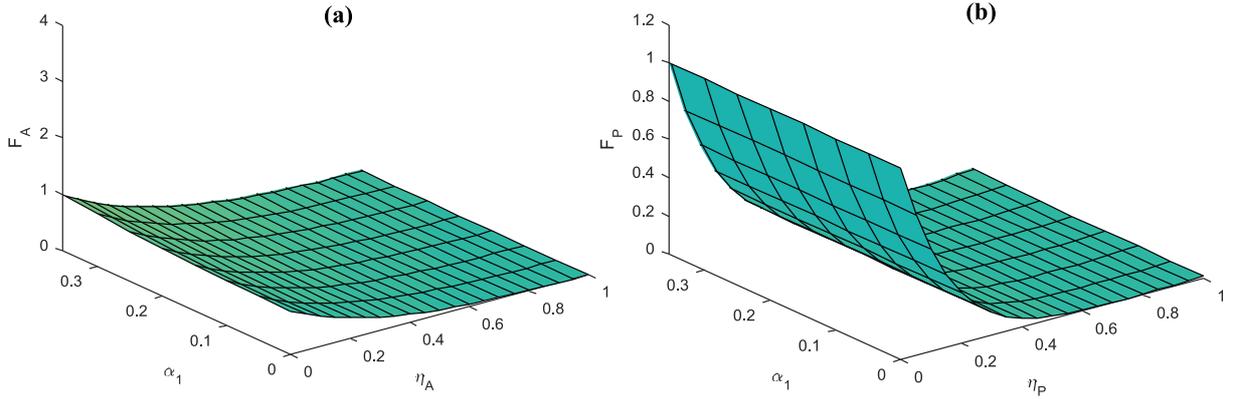}
\caption{\protect\small Three dimensional visualization of noise effect via fidelity. (a) Fidelity for amplitude-damping noise as a function of decoherence rate  and amplitude information. (b) Same as (a) but for phase-damping nose. In both cases, we have used $[\{\eta_{_A},\eta_{_P}\}, \alpha_1] = meshgrid([0:.05:1])$ and $\alpha_2=\alpha_3=...\alpha_8=1/(2\sqrt{2})$. The fidelities dwindle with increasing $\eta_{_A}$ or $\eta_{_P}$.} 
\label{fig2}
\end{figure*}
\subsection{JRSP in depolarizing channel}
In this subsection, we examine the effect of depolarizing noise on the JRSP of an arbitrary three-particle state. Depolarizing channel can be described as a model that has outstanding symmetry properties. The behavior of this noise is characterized by the following set of Kraus operators \cite{A21,A22}
\begin{equation}
E_0^D=\sqrt{1-\eta_{_D}}{\bf1},\ \ \ \ E_1^D=\sqrt{\frac{\eta_{_D}}{3}}{\bf\sigma_1},\ \ \ \ E_2^D=\sqrt{\frac{\eta_{_D}}{3}}{\bf\sigma_2}\ \ \ \mbox{and}\ \ \ \ E_3^D=\sqrt{\frac{\eta_{_D}}{3}}{\bf\sigma_3},
\end{equation}
where $\eta_{_D}(0\leq\eta_{_D}\leq 1)$ denotes the decoherence rate for the phase-damping noise. Since only qubits trios $(1,4,7)$ and $(1,4,7)$ are transmitted through noisy channel, thus the effect of the depolarizing noise on the shared entangled state can be expressed  as
\begin{equation}
\mathcal{D}(\rho)=\sum_{i,j}E_i^{D^1}\otimes E_i^{D^4}\otimes E_i^{D^7}\otimes E_j^{D^2}\otimes E_j^{D^5}\otimes E_j^{D^8}\rho\left(E_i^{D^1}\otimes E_i^{D^4}\otimes E_i^{D^7}\otimes E_j^{D^2}\otimes E_j^{D^5}\otimes E_j^{D^8}\right)^\dag,
\end{equation}
where $i,j\in\{0,1,2,3\}$ and the superscripts $(^{147258})$ denote the action of operator $E$ on which qubit. Following the same calculation of subsections $3.1$ and $3.2$, we obtain the fidelity as
\begin{eqnarray}
F=(1-{\eta}_{_D})^6+2(1-{\eta}_{_D})^3\left(\frac{{\eta}_{_D}}{3}\right)^3+\frac{{\eta}_{_D}^6}{243}.
\end{eqnarray}
The fidelity is independent of phase parameter and amplitude factor but trammel on only the decoherence rate. For $\eta_{_D}=0$, then $F=1$ which is perfect JRSP while for $\eta_{_D}=1$, $F=1/243$. Figure 3 shows the variation of fidelities as a function of decoherence rate for various noisy channel. We found that for $0.55\leq\eta\leq1$, the states transferred through depolarizing channel lose more information than through phase-damping channel.
\begin{figure*}[!t]
\centering \includegraphics[height=85mm, width=120mm]{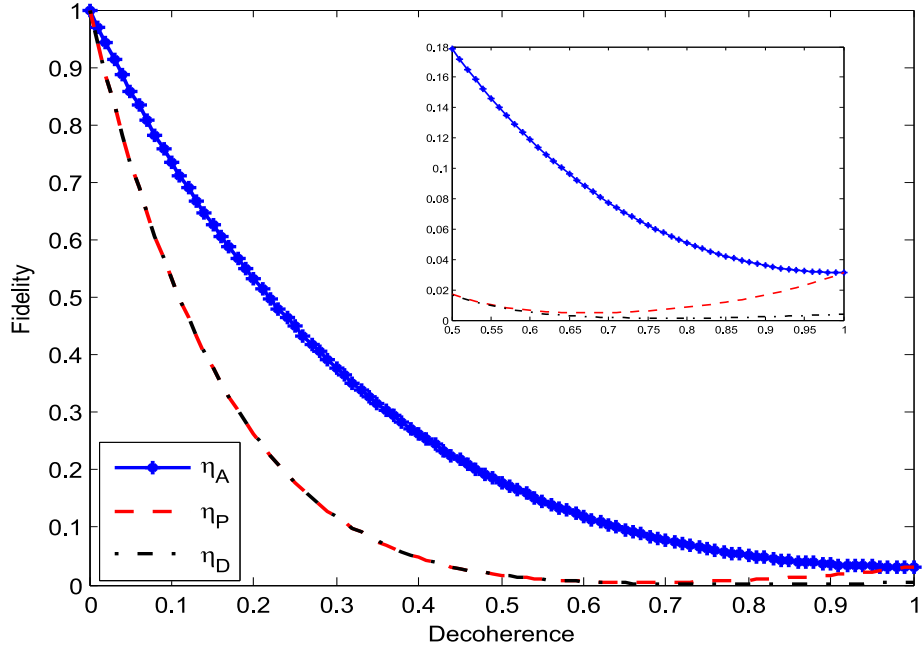}
\caption{\protect\small Plots of fidelities as a function of dechoherence rate of amplitude-damping channel, phase-damping channel and depolarizing. In all cases, we have used $\alpha_2=\alpha_3=...\alpha_8=1/(2\sqrt{2})$. The fidelities decrease with increasing $\eta_{_A}$, $\eta_{_P}$ and $\eta_{_D}$. The plot indicates that the state transmitted through depolarizing channel lose more information than phase-damping and amplitude-damping channels.}
\label{fig3}
\end{figure*}
\section{Conclusions}
This study reports scheme for JRSP of three-particle state via three tripartite GHZ class  as the quantum channel linking the three parties. Eight-qubit mutually orthogonal basis vector has been utilized as measurement basis. Alice and Bob independently performs projective measurement on their particles and then communicate the results to Ckika via classical channel. Depending on the outcome of the measurements, Chika utilize a pertinent quantum gate to realize the particle's state which Alice and Bob intend to prepare for her from the shared entangled state.  We found that the probability of success for this scheme is $1/8$ but by putting some special cases into consideration, it can be ameliorated to $1/4$ or $1$. The effect of amplitude-damping, phase-damping and depolarizing quantum noises on this scheme have been scrutinized and the analytical derivation of the fidelities for the quantum noisy channels have been presented. We found that for $0.55\leq\eta\leq1$, the states transferred through depolarizing channel lose more information than phase-damping channel while the information loss through amplitude damping channel is most minimal. 

Moreover, from the result shown in Figure 3 and the one obtained in refs. \cite{A13, A27}, we can infer that the major difference between amplitude-damping channel and phase-damping channel is that the former is more decoherent than the later. In fact, this can be seen clearly from our Figure 3, suppose we consider a particular fidelity say $F=0.3$, the corresponding $\eta_P\approx0.2$ whereas $\eta_A\approx0.35$. It is worth mentioning that, in this article, we have only consider the local decoherence. However, the Markovian master equation corresponding to nonlocal decoherence had already been discussed extensively in refs. (\cite{A24,A25,A26} and refs. therein).   

Lastly, the names (Alice, Bob and Chika) we have adopted here are for convenience. For instance, Alice and Bob jointly prepare three-qubit state for Chika is easier to follow than parties $A$ and $B$ jointly prepare three-qubit state for party $C$. This study is another boosterish evidence that justifies quantum entanglement as key resource in quantum information science.

\section*{Acknowledgments}
We thank the referees for the positive enlightening comments and suggestions, which have greatly helped us in making improvements to this paper. In addition, B.J.F. acknowledges eJDS (ICTP) and Prof. K. J. Oyewumi for his continuous supports. This work was partially supported by 20160978-SIP-IPN, Mexico.

\end{document}